\documentclass{aa_old}
\usepackage{psfig}

\def\ltsima{$\; \buildrel < \over \sim \;$}
\def\lsim{\lower.5ex\hbox{\ltsima}}
\def\gtsima{$\; \buildrel > \over \sim \;$}
\def\gsim{\lower.5ex\hbox{\gtsima}}
\def\fdeg{\hbox{$\,.\!\!^{\circ}$}}

\begin{document}

\title{High-redshift blazar identification for Swift 
J1656.3$-$3302\thanks{Partly based on X--ray observations with {\it 
INTEGRAL}, an ESA project with instruments and science data centre funded 
by ESA member states (especially the PI countries: Denmark, France, 
Germany, Italy, Switzerland, Spain), Czech Republic and Poland, and with 
the participation of Russia and the USA, and on optical observations 
collected at ESO (La Silla, Chile) under programme 079.A-0171(A).}}

\author{N. Masetti\inst{1},
E. Mason\inst{2},
R. Landi\inst{1},
P. Giommi\inst{3},
L. Bassani\inst{1},
A. Malizia\inst{1},
A.J. Bird\inst{4},
A. Bazzano\inst{5}, \\
A.J. Dean\inst{4},
N. Gehrels\inst{6},
E. Palazzi\inst{1} and
P. Ubertini\inst{5}
}

\institute{
INAF - Istituto di Astrofisica Spaziale e Fisica Cosmica di Bologna, 
via Gobetti 101, I-40129 Bologna, Italy
\and
European Southern Observatory, Alonso de Cordova 3107, Vitacura, 
Santiago, Chile
\and
ASI Science Data Center, via Galileo Galilei, I-00044 Frascati, Italy
\and
School of Physics \& Astronomy, University of Southampton, Southampton,
Hampshire, SO17 1BJ, United Kingdom
\and
INAF -- Istituto di Astrofisica Spaziale e Fisica Cosmica di
Roma, Via Fosso del Cavaliere 100, I-00133 Roma, Italy
\and
NASA/Goddard Space Flight Center, Greenbelt, MD 20771, USA
}

\titlerunning{The high-redshift blazar Swift J1656.3$-$3302}
\authorrunning{Masetti et al.}

\offprints{N. Masetti, {\tt masetti@iasfbo.inaf.it}}

\date{Received 23 October 2007; accepted 14 January 2008}

\abstract{
We report on the high-redshift blazar identification of a new gamma--ray 
source, Swift J1656.3$-$3302, detected with the BAT imager onboard the 
{\it Swift} satellite and the IBIS instrument on the {\it INTEGRAL}
satellite. Follow-up optical spectroscopy has allowed us to identify the 
counterpart as an $R\sim$ 19 mag 
source that shows broad Lyman-$\alpha$, Si {\sc iv}, He {\sc ii}, C {\sc 
iv}, and C {\sc iii}] emission lines at redshift $z$ = 2.40$\pm$0.01.
Spectral evolution is observed in X--rays when the {\it INTEGRAL}/IBIS data 
are compared to the {\it Swift}/BAT results, with the spectrum
steepening when the source gets fainter.
The 0.7--200 keV X--ray continuum, observed with {\it Swift}/XRT and 
{\it INTEGRAL}/IBIS, shows the power law shape typical of radio loud (broad 
emission line) active galactic nuclei (with a photon index $\Gamma \sim$ 1.6) 
and a hint of spectral curvature below $\sim$2 keV, possibly due to intrinsic 
absorption (N$_{\rm H} \sim$ 7$\times$10$^{22}$ cm$^{-2}$) local to the source. 
Alternatively, a slope change ($\Delta$$\Gamma$ $\sim$ 1) around 2.7 keV 
can describe the X--ray spectrum equally well. At this redshift, the observed 
20--100 keV luminosity of the source is $\sim$10$^{48}$ erg s$^{-1}$
(assuming isotropic emission), making Swift J1656.3$-$3302 one of the most 
X--ray luminous blazars. This source is yet another example 
of a distant gamma--ray loud quasar discovered above 20 keV. It is also 
the farthest object, among the previously unidentified {\it INTEGRAL} 
sources, whose nature has been determined {\it a posteriori} through 
optical spectroscopy.
}

\maketitle

\section{Introduction}

Blazars are distant and powerful active galactic nuclei (AGNs) which are 
oriented in such a way that a jet expelled from the central black hole is 
directed at small angles with respect to the observer's line of sight 
(for a recent review, see Padovani 2007). In the widely adopted scenario 
of blazars, a single population of high-energy electrons in a 
relativistic jet radiates over the entire electromagnetic spectrum via 
synchrotron and inverse Compton processes, the former dominating at low 
energies, the latter being relevant at high energies (Ghisellini et al. 
1998). The ambient photons that are inverse Compton scattered can be 
either internal (synchrotron self-Compton) and/or external 
(external Compton scattering) to the jet. As a consequence, the 
spectral energy distribution (SED) of blazars shows a double-humped 
shape, with the synchrotron component peaking anywhere from infrared to 
X--rays and the inverse Compton emission extending up to GeV/TeV 
gamma rays.

To explain the various SED shapes observed in blazars, Fossati et al. 
(1998) proposed the so-called ``blazar sequence", according to which a 
relation between peak energies and $\gamma$-dominance (the luminosity 
ratio of the second to the first peak) is present as a function of the 
source total power. This means that more luminous sources have both 
synchrotron and inverse Compton peaks located at lower energies and are 
more gamma--ray dominated than their fainter (and generally lower 
redshift) analogues.

Within the blazar population, high-redshift objects are the most 
luminous and generally belong to the class of Flat-Spectrum Radio 
Quasars (FSRQ). Observations of high-luminosity blazars in the 
X--/gamma--ray band are particularly important (especially if available 
over a broad energy range) as they allow the characterization of the 
inverse Compton peak and related parameters. More specifically, a 
flattening in the spectral distribution of the seed photons producing 
X--rays via inverse Compton is often observed at low energies 
in the X--ray spectra of these objects and can be measured 
only with broad band data (see e.g. Tavecchio et al. 2007 and 
references therein).

Unfortunately, the situation is far more complex, as absorption 
intrinsic to the source can also reproduce the spectral curvature 
observed in the X--ray band (e.g., Page et al. 2005; 
Yuan et al. 2006); in this case, information on the absorption 
is useful to understand the source environment and its relation to the 
jet. Besides this, X--/gamma--ray observations can provide evidence for 
the existence of extreme blazars, i.e. those with the synchrotron peak 
lying at X--ray energies (Giommi et al. 2007; Bassani et al. 2007). 

Here, we report detailed information on a new, powerful and hard X--ray 
selected blazar, Swift J1656.3$-$3302, recently discovered through high-energy 
observations made with {\it Swift}/BAT and {\it INTEGRAL}/IBIS. We present 
the results of our optical follow-up work, which has allowed the 
identification of the source with a blazar at redshift $z$ = 2.4, 
along with an accurate analysis of the available {\it Swift}/XRT and 
{\it INTEGRAL}/IBIS data. We also construct a SED for Swift 
J1656.3$-$3302 and discuss the characteristics of the source broad 
band emission.

The paper is structured as follows: Sect. 2 reports a collection of the 
main results available in the literature on this source; Sects. 3 and 4
illustrate the optical and high-energy observations, respectively; Sect. 
5 contains the results of this observational campaign, while a 
discussion on them is given in Sect. 6. Conclusions are 
outlined in Sect. 7. Throughout the paper, and unless otherwise 
specified, uncertainties are given at the 90\% confidence level. 
We also assume a cosmology with $H_0$ = 70 km s$^{-1}$ Mpc$^{-1}$, 
$\Omega_\Lambda$ = 0.7 and $\Omega_{\rm m}$ = 0.3.

\section{Previous information on Swift J1656.3$-$3302}

The high-energy source Swift J1656.3$-$3302 was discovered with the BAT 
imager (Barthelmy et al. 2005) onboard the {\it Swift} satellite 
(Gehrels et al. 2004) during a survey performed between December 2004 
and September 2005 (Okajima et al. 2006). The object was detected
at coordinates RA = 16$^{\rm h}$ 56$^{\rm m}$ 19$\fs$2, 
Dec = $-$33$^\circ$ 01$'$ 48$''$ (J2000), which are about 
6$\fdeg$3 from the Galactic Plane, and with a positional uncertainty of 
12$'$. Okajima et al. (2006) also reported that the 
14--200 keV BAT spectrum was very hard with a photon index 
$\Gamma$=1.3$\pm$0.3, and with a flux of 1.0$\times$10$^{-10}$ erg 
cm$^{-2}$ s$^{-1}$. The source flux was variable by a factor 
of as high as 4.

Subsequent pointed observations (Tueller et al. 2006) with {\it 
Swift}/XRT (Burrows et al. 2005) performed in June 2006 located the 
X--ray counterpart at a position RA = 16$^{\rm h}$ 56$^{\rm m}$ 
16$\fs$56, Dec= $-$33$^\circ$ 02$'$ 09$\farcs$3 (J2000), with an 
uncertainty of 
3$\farcs$7. The XRT data were fitted with an absorbed power law with 
$\Gamma$=1.4$\pm$0.3 and N$_{\rm H}$ = (3.9$\pm$0.17)$\times$10$^{21}$ 
cm$^{-2}$. By comparing this value with the Galactic absorption column 
density along 
the direction of Swift J1656.3$-$3302 (2.2$\times$10$^{21}$ cm$^{-2}$; Dickey 
\& Lockman 1990), Tueller et al. (2006) suggested that it may be an 
extragalactic object. These authors also measured an absorbed flux in the 
0.3--10 keV band of 5.6$\times$10$^{-12}$ erg cm$^{-2}$ s$^{-1}$ and 
predicted a flux in the 14--195 keV band of 4.0$\times$10$^{-11}$ erg 
cm$^{-2}$ s$^{-1}$, comparable with the BAT detection of Okajima et al. 
(2006).

Tueller et al. (2006) detected no sources with {\it Swift}/UVOT (Roming et 
al. 2005) in the XRT error circle at a limiting magnitude of 20 in the 
ultraviolet filters. The soft X--ray position of Swift J1656.3$-$3302 
is consistent with the radio source NVSS J165616$-$330211 (having a 1.4 GHz 
flux density of 410.7$\pm$12.3 mJy; Condon et al. 1998) and (albeit 
marginally) with the faint {\it ROSAT} source 1RXS J165616.6$-$330150 
(Voges et al. 2000). According to Tueller et al. (2006) the {\it ROSAT} 
data indicate that, if the two sources are the same, spectral variability 
may be present. A second, weak X--ray source was also found with XRT 
within the BAT error circle, but Tueller et al. (2006) judged it too soft 
and faint to be the soft X--ray counterpart of the BAT source.

Swift J1656.3$-$3302 was also found to be associated with
an unidentified {\it INTEGRAL} 
source in the 3$^{\rm rd}$ IBIS survey of Bird et al. (2007), with average 
20--40 keV and 40--100 keV fluxes of 9.1$\times$10$^{-12}$ erg cm$^{-2}$ 
s$^{-1}$ and 1.4$\times$10$^{-11}$ erg cm$^{-2}$ s$^{-1}$, respectively, 
assuming a Crab-like spectrum. According to these observations, the
source lies at coordinates RA = 16$^{\rm h}$ 56$^{\rm m}$ 26$\fs$4, 
Dec= $-$33$^\circ$ 02$'$ 49$\farcs$2 (J2000). The error circle is
2$\farcm$5 in radius. This position is consistent with the BAT and XRT 
positions reported above.

In the following, we further analyze this source over a broad range of 
frequencies.

\section{Optical observations}

\begin{figure}
%\begin{center}
%\hspace{.2cm}
\psfig{figure=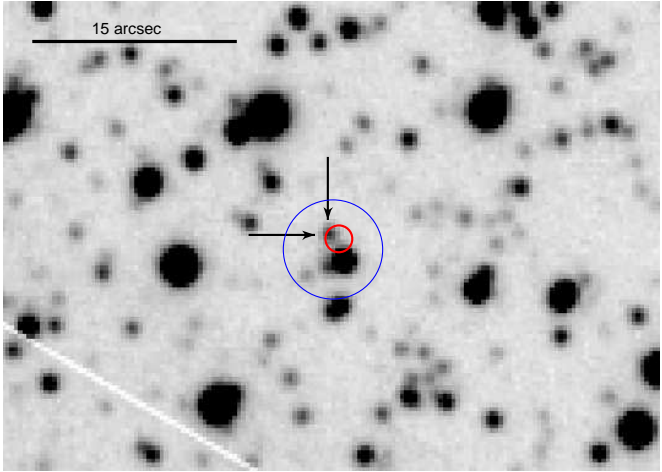,width=8.8cm,angle=0}
%\end{center}
%\vspace{-0.7cm}
\caption[]{Section of the ESO-3.6m plus EFOSC2 $R$-band acquisition 
image (exposure time: 20 s) of the field of Swift J1656.3$-$3302, with 
superimposed the 
X--ray 0.3--10 keV band {\it Swift}/XRT (larger circle) and the radio 1.4 
GHz NVSS (smaller circle) positions. The actual optical counterpart, 
identified through optical spectroscopy at the same telescope, is 
indicated by the black arrows. In the image, North is at top, 
East is to the left.}
\end{figure}

\begin{figure}%[th!]
%\begin{center}
%\vspace{-5cm}
\hspace{-.9cm}
\psfig{figure=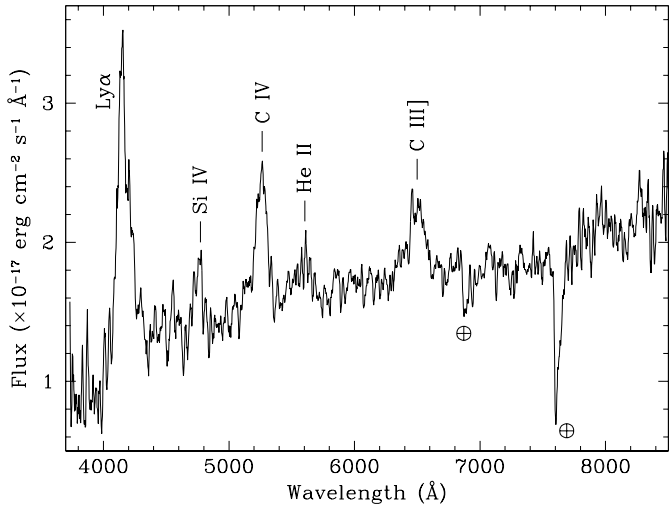,width=10.5cm,angle=-90}
%\end{center}
\vspace{-.7cm}
\caption[]{3700-8500 \AA~optical spectrum of Swift J1656.3$-$3302 
obtained with the ESO-3.6m telescope. The spectrum shows broad emission 
lines of Lyman-$\alpha$, Si {\sc iv}, He {\sc ii}, C {\sc iv} and C {\sc 
iii}] with redshift $z$ = 2.40$\pm$0.01. Telluric absorption bands are 
marked with the symbol $\oplus$. The spectrum has been smoothed with a 
running boxcar of 5 pixels ($\sim$14 \AA).}
\end{figure}

Medium-resolution optical spectra of the objects in the {\it Swift}/XRT 
error circle (see Fig. 1) were acquired between 04:24 and 06:23 UT of 21 
June 2007, and between 03:34 and 04:04 UT of 24 June 2007 with the 3.6-metre 
ESO telescope located in La Silla (Chile). This telescope carried the 
EFOSC2 instrument, equipped with a 2048$\times$2048 pixel Loral/Lesser 
CCD. The use of grating \#13 and a slit of 1$\farcs$0 provided a 
3685--9315 \AA~nominal spectral coverage. This setup gave a dispersion of 
2.8~\AA/pix.

The spectra (Fig. 2), after correction for cosmic-ray rejection, bias and 
flat-field, were optimally extracted (Horne 1986) using IRAF\footnote{IRAF 
is the Image Analysis and Reduction Facility made available to the 
astronomical community by the National Optical Astronomy Observatories, 
which are operated by AURA, Inc., under contract with the U.S. National 
Science Foundation. It is available at {\tt http://iraf.noao.edu/}}. 
Wavelength calibration was performed using He-Ar lamps, while flux 
calibration was accomplished by using the spectrophotometric standard 
Feige 110 (Hamuy et al. 1992, 1994). The wavelength calibration uncertainty 
was $\sim$0.5~\AA; this was checked by using the positions of background 
night sky lines.

A 20 s $R$-band acquisition image of the field, secured
again with the 3.6m ESO telescope plus EFOSC2 under a seeing of 
0$\farcs$9, was also reduced and analyzed. The image, 2$\times$2 
pixels binned, had a scale of 0$\farcs$31/pix and covered a field of 
5$\farcm$2$\times$5$\farcm$2. It was corrected for bias and flat field and 
processed with {\sc daophot} (Stetson 1987) within 
MIDAS\footnote{MIDAS (Munich Image Data Analysis System) is developed, 
distributed and maintained by ESO and is available at {\tt 
http://www.eso.org/projects/esomidas/}} using a PSF-fitting procedure. The 
choice of this photometric approach over the simple aperture photometry 
was dictated by the crowdedness of the field (see Fig. 1).

\section{X--/gamma--ray observations}

The field of Swift J1656.3$-$3302 was observed twice with XRT onboard 
{\it Swift} in the 0.3--10 keV range (see Fig. 3). The first 
observation started at 03:32 UT on 9 June 2006, and the second at 
16:36 UT on 13 June 2006. The pointings had
on-source times of 4.4 and 4.8 ks, respectively, and both were 
performed in Photon Counting mode (see Burrows et al. 2005 for details on 
this observing mode).

\begin{figure}
\psfig{figure=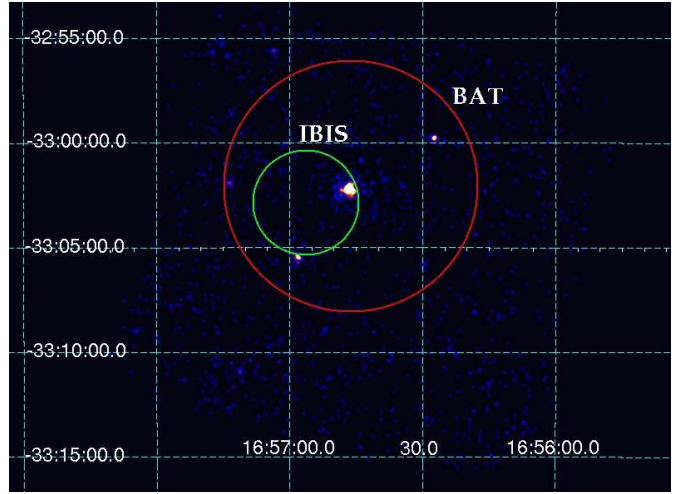,width=8.8cm,angle=0}
\caption[]{{\it Swift}/XRT 0.3--10 keV image of the field of Swift 
J1656.3$-$3302 acquired on 13 June 2006. The larger circle indicates
the BAT positional uncertainty and the smaller circle the 
IBIS one. Swift J1656.3$-$3302 is the brightest source in the image.}
\end{figure}

Data reduction was performed using the XRTDAS v2.0.1 standard data 
pipeline package ({\sc xrtpipeline} v10.0.6). Events for spectral 
analysis were extracted within a circular region of radius 20$''$ (which 
encloses about 90\% of the PSF at 1.5 keV; Moretti et al. 2004) centered 
on the source position. The background was extracted from a circular 
region located far from the source. In all cases, the spectra were 
extracted from the corresponding event files using {\sc xselect} software 
and binned using {\sc grppha}, so that the $\chi^{2}$ statistic could 
reliably be used. We used version v.008 of the response matrices in the 
Calibration Database\footnote{available at: {\tt 
http://heasarc.gsfc.nasa.gov/\\docs/heasarc/caldb/caldb\_intro.html}} 
(CALDB 2.3) maintained by HEASARC and we created individual ancillary 
response files using {\sc xrtmkarf} v.0.5.2 within 
FTOOLS\footnote{available at:\\ 
\texttt{http://heasarc.gsfc.nasa.gov/ftools/}} (Blackburn 1995). The 
X--ray spectral analysis was performed with the package {\sc xspec} 
v.11.3.2 (Arnaud 1996).

We also extracted the spectral data of this source collected with the 
coded-mask ISGRI detector (Lebrun et al. 2003) of the IBIS instrument 
(Ubertini et al. 2003) onboard {\it INTEGRAL} (Winkler et al. 2003). 
ISGRI data were processed using the standard {\it INTEGRAL} analysis 
software (OSA\footnote{available at: \\ {\tt 
http://isdc.unige.ch/index.cgi?Soft+download}} v5.1; Goldwurm et al. 
2003). Events in the band 20--200 keV, coming from both fully-coded and 
partially-coded observations of the field of view of Swift 
J1656.3$-$3302, were included in the analysis. Details on the whole 
extraction procedure can be found in Bird et al. (2007). A time-averaged 
spectrum was obtained from the available data using the method 
described in Bird et al. (2006, 2007); that is, the spectra were 
reconstructed using the information obtained from the flux maps acquired 
in several bands between 20 and 200 keV. Data for a total exposure of 
2.3 Ms, collected in the time interval October 2002 - April 2006, were 
used for this task.

\section{Results}

\begin{figure}
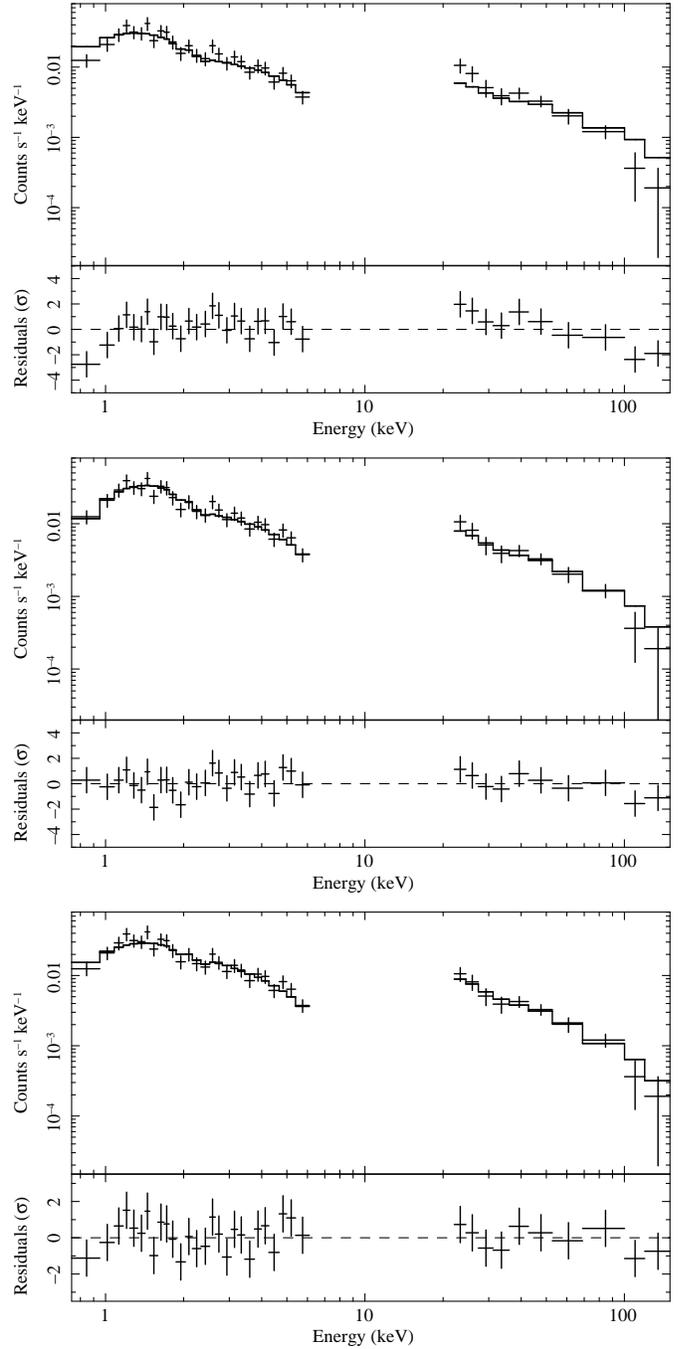

%\begin{center}
%\hspace{.2cm}
\psfig{figure=plot_wagal_zpo_bw.ps,width=8.8cm,angle=-90}
\vspace{0.2cm}
\psfig{figure=plot_wagal_zwa_zpo_bw.ps,width=8.8cm,angle=-90}
\vspace{0.2cm}
\psfig{figure=plot_wagal_bknpower_bw.ps,width=8.8cm,angle=-90}
%\end{center}
\vspace{-0.1cm}
\caption[]{Averaged 0.7--200 keV X--ray spectrum of Swift J1656.3$-$3302
obtained from the XRT and ISGRI data described in the text and fitted 
with a power law absorbed only by the Galactic hydrogen column 
({\it upper panel}), a power law absorbed by Galactic plus neutral 
hydrogen local to the blazar ({\it central panel}), and a broken power 
law absorbed by the Galactic hydrogen column only ({\it lower panel}). 
The fit residuals using the best-fit models reported in Table 2 are 
shown in each panel.}
\end{figure}

The 0.3--10 keV image of the region around Swift J1656.3$-$3302 is shown 
in Fig. 3 with the IBIS and BAT error circles superimposed. Three 
objects are clearly detected within the BAT error circle. However, 
only one, which is also the brightest, lies within the smaller {\it INTEGRAL} 
error circle. This object is the soft X--ray counterpart of Swift 
J1656.3$-$3302 reported in Tueller et al. (2006).

With the data of both XRT observations we have determined its position 
using the most recent version of the {\tt xrtcentroid} (v0.2.7) task. The 
correction for the misalignment between the telescope and the satellite 
optical axis was taken into account (see Moretti et al. 2006 for 
details). The position we obtained for the source is 
RA = 16$^{\rm h}$ 56$^{\rm m}$ 16$\fs$83, Dec = $-$33$^\circ$ 
02$'$ 12$\farcs$3 (J2000), with an uncertainty of 3$\farcs$7 on 
both coordinates. This is fully consistent with the preliminary 
results of Tueller et al. (2006).

As mentioned above, in addition to the bright X--ray source,
two fainter sources are detected by XRT (see Fig. 3).
One of them (the westernmost one) can readily be excluded as 
the soft X--ray counterpart of Swift J1656.3$-$3302 because it is 
positionally inconsistent with the IBIS error circle.
The other lies 2$\farcm$62 from the center of the IBIS error 
circle, near its border but formally outside it. 
This object has coordinates RA = 16$^{\rm h}$ 56$^{\rm m}$ 
28$\fs$1, Dec= $-$33$^\circ$ 05$'$ 25$\farcs$2 (J2000), with a 
conservative error of 6$''$ on both. To further exclude any 
association between this source 
and the hard X--ray emission seen by BAT and IBIS, we examined its XRT 
spectrum. We found that it is very soft, as no emission is detected above 
2.5 keV. Indeed, it could be fitted with a blackbody with $kT$ = 
0.5$^{+0.8}_{-0.2}$ keV ($\chi^2$/dof = 3.8/3). The corresponding 2--10 keV 
flux is 1.3$\times$10$^{-13}$ erg cm$^{-2}$ s$^{-1}$, thus (see Table 
2) it is about 30 times fainter than the aforementioned bright 
XRT source, which is located within the {\it INTEGRAL} error circle.

The positional coincidence of this faint X--ray source with an
$R \sim$ 11.4 mag star in the DSS-II-red survey suggests that this soft 
X--ray emission comes from a stellar corona. Indeed, medium-resolution 
optical spectroscopy acquired at La Silla (Chile) with the NTT 
equipped with EMMI on 2007 July 28 shows that this source is an A-type 
star with no peculiar spectral features.

Given all of the above multiwavelength information, we can confidently 
exclude the two fainter objects, detected by XRT and shown in Fig. 3, as 
the possible soft X--ray counterparts of Swift J1656.3$-$3302, and we 
confirm that the object reported in Tueller et al. (2006) is the soft 
X--ray counterpart of the source detected by BAT and IBIS.

Of the 5 optical objects within or close to the XRT error circle
of Swift J1656.3$-$3302
(see Fig. 1), only that indicated with the tick marks in this figure 
shows peculiar optical spectroscopic characteristics, namely broad emission 
lines (Fig. 2). We identify these features as Lyman-$\alpha$, Si {\sc 
iv} $\lambda\lambda$1394,1403, C {\sc iv} $\lambda\lambda$1548,1551, He 
{\sc ii} $\lambda$1640 and C {\sc iii}] $\lambda$1910, at an average 
redshift $z$ = 2.40$\pm$0.01. This corresponds to a luminosity distance 
$d_L$ = 19.4 Gpc within the assumed cosmology. Table 1 reports the 
main properties of the emission lines detected in the optical spectrum of
Swift J1656.3$-$3302. 

We note that the Lyman-$\alpha$, He {\sc ii} and C {\sc iii}] 
emission lines have much broader Full Widths at Half Maximum (FWHMs) than 
the other lines. Given the resolution and the signal-to-noise ratio of the 
optical spectrum, this may be due to the fact that these lines may be 
blended with the N {\sc v} $\lambda$1240, O {\sc iii}] $\lambda$1663 and 
Si {\sc iii}] $\lambda$1892 emissions, respectively.

The $R$-band acquisition image was processed to obtain an astrometric 
solution based on several USNO-A2.0\footnote{The USNO-A2.0 catalogue is 
available at \\ {\tt http://archive.eso.org/skycat/servers/usnoa}} 
reference stars in the field of Swift J1656.3$-$3302. This yields for the 
optical counterpart of this source the coordinates RA = 16$^{\rm 
h}$ 56$^{\rm m}$ 16$\fs$853, Dec = $-$33$^{\circ}$ 02$'$ 11$\farcs$08
(J2000). The conservative error on the optical position is 
0$\farcs$31, which has to be 
added to the systematic error of the USNO catalogue (0$\farcs$25 according 
to Assafin et al. 2001 and Deutsch 1999). The final 1-$\sigma$ astrometric 
uncertainty on the optical position of Swift J1656.3$-$3302 is thus 
0$\farcs$40.

This position is consistent with that of the radio source NVSS 
J165616$-$330211, indicating that the two sources are the same. 
This is also consistent with the position of the
faint {\it ROSAT} source reported by Tueller et 
al. (2006). We instead exclude that the optical source USNO-A2.0 
0525-24886745 (i.e. the brightest object within the XRT error circle in 
Fig. 1, proposed by Tueller et al. 2006 as a possible optical 
counterpart) is related with Swift J1656.3$-$3302 on the basis of its 
optical spectrum, which is typical of a Galactic star.

\begin{table}[h!]
\caption[]{Main properties of the emission lines detected in the optical 
spectrum of Swift J1656.3$-$3302. Line fluxes are reported with and 
without correction for the intervening Galactic extinction (see text).
Equivalent Widths (EWs) are expressed in the observer's frame.
Uncertainties are at 1-$\sigma$ confidence level.}
%\vspace{-.4cm}
\begin{center}
\begin{tabular}{lcccc}
\noalign{\smallskip}
\hline
\hline
\noalign{\smallskip}

\multicolumn{1}{c}{Line} & \multicolumn{2}{c}{Flux (10$^{-15}$ erg cm$^{-2}$ s$^{-1}$)} & EW & FWHM \\
\cline{2-3}
\noalign{\smallskip}
 & observed & extinction- & (\AA) & (km s$^{-1}$) \\
 & & corrected & & \\

\noalign{\smallskip}
\hline
\noalign{\smallskip}

Lyman-$\alpha$$^*$ & 2.8$\pm$0.3   & 39$\pm$4    & 263$\pm$26 & $\sim$9900  \\
Si {\sc iv}        & 0.41$\pm$0.12 & 4.4$\pm$1.3 &  30$\pm$10 & $\sim$5000  \\
C {\sc iv}         & 1.19$\pm$0.12 & 7.2$\pm$0.7 &  80$\pm$8  & $\sim$6600  \\
He {\sc ii}$^*$    & 0.66$\pm$0.13 & 2.3$\pm$0.5 &  43$\pm$9  & $\sim$12000 \\
C {\sc iii}]$^*$   & 1.1$\pm$0.1   & 5.1$\pm$0.5 &  65$\pm$7  & $\sim$9300  \\

\noalign{\smallskip}
\hline
\noalign{\smallskip}
\multicolumn{5}{l}{$^*$: possibly blended with a fainter emission 
line (see text)} \\
\noalign{\smallskip}
\hline
\hline
\end{tabular}
\end{center}
\end{table}

Again using USNO-A2.0 field stars as calibrators, we obtained for the 
true counterpart a magnitude $R$ = 19.1$\pm$0.1. Towards the line of 
sight of the source, the Galactic foreground reddening is $E(B-V)$ = 
0.624 mag (Schlegel et al. 1998). Using the law by Cardelli et al. 
(1989), this implies that the Galactic extinction in the $R$ band is 
$A_R$ = 1.6. Thus, the dereddened $R$-band magnitude of the 
source is $R_0$ = 17.5.

A simple power law model (as employed by Okajima et al. 2006 to 
characterize the BAT spectrum) provides a good description of the IBIS 
data but yields a steeper spectrum ($\Gamma$ = 1.9$\pm$0.3), combined 
with a lower (by a factor of 3 compared to the BAT measurement) 
14--195 keV flux (3.4$\times$10$^{-11}$ erg cm$^{-2}$ s$^{-1}$). 
Thus, the source appears to experience spectral evolution. 

We then used the {\it Swift}/XRT data to describe the X--ray 
spectrum of Swift J1656.3$-$3302, employing first a simple power law 
in the source rest frame ({\tt zpowerlw} model in {\sc xspec}) 
absorbed by the Galactic column density (reported in Sect. 2).
This model provides consistent results for the two observations 
($\Gamma$=1.22$\pm$0.16 for the first one and $\Gamma$=1.13$\pm$0.16 for 
the second). No significant flux variations are observed either 
during each XRT pointing or between them. Thus, to improve the 
statistics, we have stacked the two XRT spectra and repeated the 
analysis. The simple power law again provides an acceptable fit
($\chi^{2}$/dof = 24.8/25) and yields a flat spectrum 
($\Gamma$=1.18$\pm$0.11).

Next, we have combined the XRT spectrum with that from IBIS, 
introducing a 
constant to allow for intercalibration differences between the two 
instruments; this constant was left free to vary in the fits. 
We are aware that, since the XRT and IBIS observations were not 
simultaneous, flux variations are plausible given the blazar nature 
of the object, and may also be the cause of different normalizations.
Given that our past experience tells us that the intercalibration
constant between these two instruments is $\sim$1, we suggest that
different values for this constant are likely to be due to the flux 
variability mentioned above.

Again using an absorbed power law fit with the N$_{\rm H}$ value 
fixed to the Galactic one along the source line of sight, we obtained 
a best-fit photon index 
$\Gamma$=1.28$\pm$0.10 ($\chi^{2}$/dof = 45.6/34). However, this model 
results in fit residuals (see Fig. 4, upper panel) that show some 
curvature on the lower energy side, possibly due to intrinsic absorption 
in the source rest frame. The addition of this extra absorption component 
({\tt zwabs} in {\sc xspec}) provides a fit improvement which is 
significant at the 95\% confidence level according to the F-test 
(Bevington 1969). This leads to a more typical AGN spectrum ($\Gamma$= 
1.64$\pm$0.16) and gives a relatively high column density (N$_{\rm 
H}$=6.7$^{+3.2}_{-2.7}$$\times$10$^{22}$ cm$^{-2}$) local to the blazar 
(Fig. 4, central panel).

Alternatively, one may assume that the turnover in the X--ray spectrum is 
produced by an intrinsic slope change: in this case, by fitting the XRT 
and IBIS data using a broken power law absorbed by the Galactic hydrogen 
column, we find a spectral steepening $\Delta$$\Gamma$ = 0.95 occurring at 
energy $E_{\rm break}$ = 2.7 keV (Fig. 4, lower panel). Again using the 
F-test, we find that the improvement significance of this spectral 
description over the simple power law with Galactic absorption is 95\% 
in this case also.

The XRT+IBIS global fit results for the three models above are reported in 
Table 2.
With the assumed cosmology and using the redshift of the source, we get 
2--10 keV and 20--100 keV observer's frame luminosities of 
2.1$\times$10$^{47}$ erg s$^{-1}$ and 9.9$\times$10$^{47}$ erg s$^{-1}$
respectively, assuming isotropic emission. These figures are largely 
independent of the assumed spectral model.

\begin{table}
\caption[]{Best-fit parameters of the models adopted to describe the 
X--ray spectrum of Swift J1656.3$-$3302.}
\vspace{-.2cm}
\begin{center}
\begin{tabular}{lr}
\noalign{\smallskip}
\hline
\hline
\noalign{\smallskip}
\multicolumn{1}{c}{Parameter} & \multicolumn{1}{c}{Value} \\
\noalign{\smallskip}
\hline
\noalign{\smallskip}
\multicolumn{2}{c}{power law plus Galactic absorption} \\
\noalign{\smallskip}
\hline
\noalign{\smallskip}
$\chi^2$/dof       & 45.6/34 \\
$\Gamma$         & 1.28$\pm$0.10 \\
XRT/IBIS intercalibr. constant    & 0.7$^{+0.4}_{-0.2}$ \\
$F_{\rm (2-10 \,keV)}$       & 4.9$\times$10$^{-12}$ \\
$F_{\rm (20-100 \,keV)}$  & 1.9$\times$10$^{-11}$ \\
\noalign{\smallskip}
\hline
\noalign{\smallskip}
\multicolumn{2}{c}{power law plus Galactic and intrinsic absorption} \\
\noalign{\smallskip}
\hline
\noalign{\smallskip}
$\chi^2$/dof & 25.2/33 \\
N$_{\rm H}$ (10$^{22}$ cm$^{-2}$) & 6.7$^{+3.2}_{-2.7}$ \\
$\Gamma$ & 1.64$\pm$0.16 \\
XRT/IBIS intercalibr. constant & 1.9$^{+1.3}_{-0.7}$ \\
$F_{\rm (2-10 \,keV)}$    & 4.6$\times$10$^{-12}$ \\
$F_{\rm (20-100 \,keV)}$  & 2.2$\times$10$^{-11}$ \\
\noalign{\smallskip}
\hline
\noalign{\smallskip}
\multicolumn{2}{c}{broken power law plus Galactic absorption} \\
\noalign{\smallskip}
\hline
\noalign{\smallskip}
$\chi^2$/dof & 23.6/32 \\
$\Gamma_1$ & 0.86$^{+0.23}_{-0.26}$ \\
$\Gamma_2$ & 1.81$^{+0.23}_{-0.13}$ \\
$E_{\rm break}$ (keV) & 2.7$^{+1.5}_{-0.5}$ \\
XRT/IBIS intercalibr. constant & 2.9$^{+1.5}_{-1.3}$ \\
$F_{\rm (2-10 \,keV)}$    & 4.4$\times$10$^{-12}$ \\
$F_{\rm (20-100 \,keV)}$  & 2.0$\times$10$^{-11}$ \\
\noalign{\smallskip}
\hline
\noalign{\smallskip}
\multicolumn{2}{l}{Note: In the above fits, when needed, we fixed the redshift} \\
\multicolumn{2}{l}{of the source at $z$=2.40 and the Galactic hydrogen column} \\
\multicolumn{2}{l}{at N$_{\rm H}^{\rm Gal}$ = 2.2$\times$10$^{21}$ cm$^{-2}$.} \\
\multicolumn{2}{l}{The reported fluxes (in the observer's frame) are in erg} \\
\multicolumn{2}{l}{cm$^{-2}$ s$^{-1}$ and are corrected for the total intervening} \\
\multicolumn{2}{l}{absorption column density.} \\
\multicolumn{2}{l}{For the broken power law model, the fit parameters} \\ 
\multicolumn{2}{l}{are computed in the observer's frame, and not in the} \\
\multicolumn{2}{l}{source rest frame.} \\
\noalign{\smallskip}
\hline
\hline
\end{tabular}
\end{center}
\end{table}

\section{Discussion}

\begin{figure}
%\begin{center}
%\hspace{.2cm}
\psfig{figure=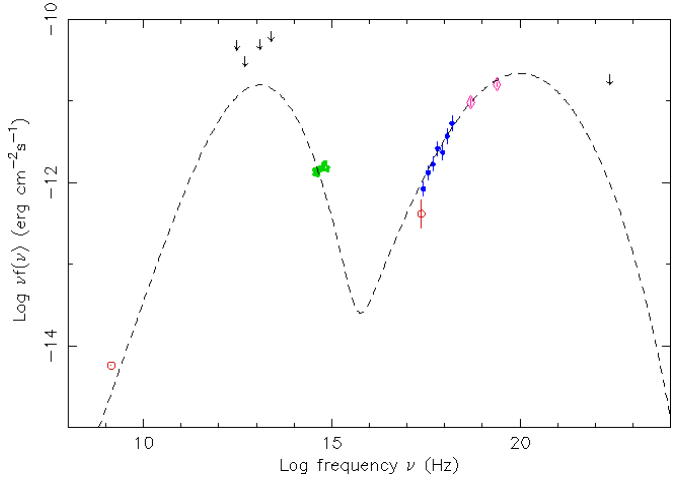,width=8.8cm,angle=0}
\vspace{-0.1cm}
\caption[]{Broadband non-simultaneous observational radio to gamma--ray 
SED of Swift J1656.3$-$3302, constructed with the measurements collected 
from the literature and with those presented in this paper
(see text for details). Moving towards 
higher energies, the following information was used: the 1.4 GHz NVSS radio 
flux, the ESO 3.6m optical spectrum, the {\it ROSAT} 0.1--2.4 keV flux, and 
the XRT+IBIS 0.7--200 keV spectrum. The {\it IRAS} upper limits in the 
mid-infrared bands and the upper limit at MeV energies from the 
EGRET survey are also reported. The dashed line indicates an example of 
the application of the log-parabolic model (Massaro et al. 2006; Tramacere 
et al. 2007) briefly described in the text.}
\end{figure}

Within our optical spectroscopy program aimed at the identification 
of the nature of sources detected at high energies with {\it INTEGRAL}
(see Masetti et al. 2006 and references therein), 
we discovered that the object Swift J1656.3$-$3302 is a high-redshift 
blazar, located at $z$ = 2.40. This source is thus another case belonging 
to the small but growing class of distant gamma--ray emitting blazars, 
which are being discovered by {\it Swift}/BAT and {\it INTEGRAL}/IBIS 
(Sambruna et al. 2006; Sambruna et al. 2007; Bassani et al. 2007; Bird et 
al. 2007). Swift J1656.3$-$3302 is moreover the farthest object, among 
the previously unidentified {\it INTEGRAL} sources, whose nature has been
determined {\it a posteriori} through optical spectroscopy.

The better optical position has allowed us to associate it with a fairly bright 
NVSS radio source and with a faint {\it ROSAT} object. The NVSS image of 
the source shows that it is core-dominated with no extended radio features.  
The optical $R$-band information reported in Sect. 4, combined with 
the radio flux density, suggests that Log($S_{\rm 1.4~GHz}/S_R$) $\sim$ 3.3 
for this object. Although this ratio is not conventionally used to 
characterize the radio loudness of a source, it can nevertheless be used 
to quantify the strength of the radio emission compared to the optical.
In this estimate we did not consider any contribution from the
optical extinction local to the source: to take this 
issue into account, we can assume for Swift J1656.3$-$3302 the lower 
extreme in the range of the optical-to-X--ray slopes typical for this kind 
of object (e.g., Stocke et al. 1991). This gives a strict upper limit 
to the $R$-band flux of the source, and hence a lower limit for the 
radio-optical flux density ratio: with this procedure, we obtain that
Log($S_{\rm 1.4~GHz}/S_R$) $>$ 2.5. In summary, the considerations above 
point to the fact that Swift J1656.3$-$3302 is very likely to be 
radio loud.

If we compare the source unabsorbed monochromatic X--ray flux at 1 
keV to the radio flux at 1.4 GHz, we find that their ratio is around 
150 or, following the notation of Giommi et al. (2007), 
$\sim$2$\times$10$^{-12}$ erg cm$^{-2}$ s$^{-1}$ 
Jy$^{-1}$, i.e. similar to blazars with the inverse Compton peak in the 
gamma--ray band. The source is indeed very powerful at high energies, 
which explains the detection by BAT and IBIS.
The above properties suggest that Swift J1656.3$-$3302 is a 
distant blazar in which the emission is relativistically beamed and with 
a double-peaked SED.

To verify this, we constructed the non-simultaneous SED of Swift 
J1656.3$-$3302 by combining all the available data in order to cover as 
many frequencies as possible (see Fig. 5). Along with the optical and 
X--/gamma--ray information obtained in the present work, we have used 
the 1.4 GHz NVSS radio flux from Condon et al. (1998), the {\it IRAS} 
infrared upper limits (IRAS 1988) in the mid-infrared bands (12, 25, 60 
and 100 $\mu$m), the {\it ROSAT} 0.1--2.4 keV flux (Voges et al. 2000) 
and the upper limit at MeV energies from the EGRET survey (Hartman et al. 
1999). When plotting the SED, we respectively corrected the optical and 
the X--ray spectra from the contribution of the foreground Galactic 
extinction using the $E(B-V)$ and N$_{\rm H}$ values reported in 
Sects. 5 and 2, respectively.
Moreover, the optical spectrum was rebinned at $\sim$60 \AA~to 
smooth out the noise of the continuum, and the most evident 
emission lines were removed.

Looking at the SED of this object, two things are immediately clear:
(i) it is indeed double peaked as expected from a blazar; and
(ii) it is typical of a powerful blazar with the synchrotron 
peak somewhere in the infrared range and that of the inverse 
Compton component just above the ISGRI band (20--100 keV). 
Thus, given the source luminosity, the optical/radio information and 
the overall SED, it is likely that Swift J1656.3$-$3302 belongs to 
the class of FSRQs.

An interesting point to recall here is the fact that, as mentioned in 
Sect. 5, the high-energy spectrum of this source shows a deficit of 
soft X--ray photons. This could be produced by either an 
intrinsically curved shape of the spectrum itself, or absorption 
local to the AGN. This feature is becoming  common place, 
rather than the exception, for high redshift blazars and it 
is a widely debated issue at the moment. Observations of intrinsic 
absorption in high redshift blazars have been reported for many objects 
(Page et al. 2005; Yuan et al. 2006 and references therein): the common 
interpretation is that this absorption originates from the material present 
in the AGN environment.

From the above mentioned literature there is some evidence of a 
correlation between absorption and redshift, with the more distant sources 
being more absorbed. The increase in N$_{\rm H}$ with $z$ seems to occur 
starting at $z\sim$ 2. Moreover, at this redshift one sees a change 
in the fraction of radio loud quasars showing X--ray absorption, being
lower at lower $z$. Finally, there is a tendency for objects with 
intrinsic absorption to have systematically higher X--ray fluxes. All this 
evidence suggests that there may be a strong cosmic evolution effect that 
takes place at $z \sim$ 2.

Within this scenario Swift J1656.3$-$3302 is quite interesting, because it 
has a redshift larger than 2 and also because it has a relatively high column 
density compared to the objects discussed by Yuan et al. (2006, see their 
Fig. 6). 

The absorption scenario is however far from being confirmed: the fact that 
absorption is more often seen in radio loud objects, i.e. those which more 
likely host a relativistic jet, raises the question of why these jets are 
not able to remove the gas in their vicinity. Another problem is the 
apparent absence of extinction at lower energies (that is, UV and optical), 
if absorption is indeed present in these objects, and in particular in
Swift J1656.3$-$3302: indeed, the shape of its optical spectrum, once 
corrected for the Galactic extinction, suggests a local column 
density which is much lower than that inferred from the X--ray data 
analysis (this value would imply a $V$-band rest-frame extinction 
$A_V \sim$ 40 mag, assuming the Galactic extinction law). This 
effect may however be partly alleviated by the fact that, at high redshift, 
the chemical composition of the dust is substantially different from that 
of the Milky Way, thus producing an optical-UV extinction law radically 
at variance with that of our Galaxy (e.g., Calzetti et al. 1998; 
Maiolino et al. 2001).

An alternative interpretation for the deficit 
of soft photons has been put forward for blazars, i.e. that the observed 
shape is due to intrinsic curvature of the inverse Compton emission
(Fabian et al. 2001; see also Tavecchio et al. 2007 for a recent 
investigation made using this model).
Indeed, a low energy cutoff in the relativistic particle distribution at 
$\gamma_{\rm cut}$ would produce a flattening in the scattered 
spectrum below $\nu_{\rm obs}$ $\sim$ $\nu_{\rm ex} \cdot \Gamma_{\rm 
L}^2 \cdot \gamma_{\rm cut}^2 \cdot (z + 1)^{-1}$, where 
$\Gamma_{\rm L} $ is the Lorentz factor, $\nu_{\rm ex}$ the frequency of 
the seed photons and $z$ the redshift of the source. 
If $\nu_{\rm ex}$ is $\approx$10$^{15}$ Hz, the break observed 
in Swift J1656.3$-$3302 implies, for Lorentz factors in the range 10--50,
that $\gamma_{\rm cut} \approx$ a few.

A different description of the source SED, again in the assumption of 
the intrinsic curvature hypothesis, can be made assuming a 
log-parabolic distribution for the emitting electrons, in the form
$F$($E$) = $K \cdot E^{-(a+b \cdot Log\,E)}$ (Massaro et al. 2006; 
Tramacere et al. 2007), where $a$ is the power law slope at 1 keV 
and $b$ is the curvature below this energy. As an example, in Fig. 5 we 
overplotted on the SED a log-parabolic distribution with $a$ = 0.55 and 
$b$ = 0.6. We however defer the detailed analysis of the intrinsic 
curvature of the X--ray spectrum of Swift J1656.3$-$3302 to a subsequent
paper.

We conclude this Section by briefly commenting on the spectral evolution 
seen when comparing the IBIS and the BAT spectra: apparently, 
the source X--ray spectrum steepens when it gets fainter. 
The observed X--ray variability in both flux and spectrum reported 
here for Swift J1656.3-3302 is unusual, though not unprecedented in 
FSRQs. Indeed, a number of other high-$z$ blazars 
show the same behavior, such as RX J1028.6$-$0844 ($z$ = 4.2; 
Yuan et al. 2005) and GB B1428+4217 ($z$ = 4.7; Worsley et al. 2006)
and, in more recent studies, RBS 315 ($z$ = 2.7; Tavecchio et al. 2007) 
and QSO 0836+710 ($z$ = 2.2; Sambruna et al. 2007). This variability 
could be due to changes in the electron distribution at low energies, 
or may be a hint of the presence of an extra component, which becomes 
stronger or weaker as the source brightness changes.

\section{Conclusions}

Through optical follow-up observations we have been able to identify the 
newly discovered hard X--ray source Swift J1656.3$-$3302 with a blazar 
at $z$ = 2.4. Spectral evolution is observed when comparing the Swift/BAT 
and {\it INTEGRAL}/IBIS data, with the X--ray spectrum undergoing softening 
when the emission becomes fainter. The source broadband X--/gamma--ray 
spectrum is well described by a power law of index $\Gamma \sim$ 1.6. 
The source is extremely bright with a 20--100 keV observer's frame 
luminosity of $\sim$10$^{48}$ erg s$^{-1}$, assuming isotropic emission. 

The source SED is typical of high luminosity blazars, as it has two peaks: 
the lower (synchrotron) one likely located at infrared frequencies, and the 
higher (inverse Compton) one positioned above a few hundred keV. 
The spectral curvature detected in the X--ray spectrum of the 
object can either be intrinsic and due to the distribution of the emitting 
electrons, or associated with the presence of absorption local to 
the source and produced by a column density of a $\sim$7$\times$10$^{22}$ 
cm$^{-2}$, higher than that typically observed in high-$z$ blazars.

All of the observational evidence gathered so far therefore points to Swift 
J1656.3$-$3302 being another case of a distant luminous blazar selected at 
gamma--ray energies. It is also the farthest object, among the 
previously unidentified {\it INTEGRAL} sources, whose nature has been
determined {\it a posteriori} through optical spectroscopy.

\begin{acknowledgements}

We thank A. S\'anchez for night assistance at the 3.6m ESO telescope,
A. Ederoclite for help and useful advices with the ESO telescopes, and
J.B. Stephen for helpful discussions. An anonymous referee 
is acknowledged for comments which helped us to improve this paper.
This research has made use of the NASA's Astrophysics Data System, of the 
HEASARC archive, and of the SIMBAD database operated at CDS, Strasbourg, 
France. The authors acknowledge the ASI and INAF financial support via 
ASI-INAF grants I/023/05/0 and I/088/07/0.

\end{acknowledgements}

\end{document}